\documentclass[10pt,tightenlines,eqsecnum,floats,aps,amsmath,
amssymb,nofootinbib,prd,noshowpacs,notitlepage,longbibliography]{revtex4-1}
\usepackage[utf8]{inputenc}
\usepackage{amsmath,amsthm,latexsym,amssymb,amsfonts,color}
\usepackage{hyperref}

\begin{document}

\title{The Barrett-Crane model: asymptotic measure factor}

\author{Wojciech Kami\'nski${}^{1,2}$}
\email{wkaminsk@fuw.edu.pl}
\author{Sebastian Steinhaus${}^{2}$}
\email{steinhaus.sebastian@gmail.com}

\affiliation{
  ${}^{1}$Instytut Fizyki Teoretycznej, Uniwersytet Warszawski,
  ul. Ho\.{z}a 69, 00-681 Warszawa, Poland\\
  ${}^{2}$Perimeter Institute for Theoretical Physics, 31 Caroline Street North
  Waterloo, Ontario Canada N2L 2Y5}

\begin{abstract}
The original spin foam model construction for $4$D gravity by Barrett and Crane suffers from a few troubling issues. In the simple examples of the vertex amplitude they can be summarized as the existence of contributions to the asymptotics from non geometric configurations. Even restricted to geometric contributions the amplitude is not completely worked out. While the phase is known to be the Regge action, the so called measure factor has remained mysterious for a decade. In the toy model case of the $6j$ symbol this measure factor has a nice geometric interpretation of $V^{-1/2}$ leading to speculations that a similar interpretation should be possible also in the $4$D case. In this paper we provide the first geometric interpretation of the geometric part of the asymptotic for the spin foam consisting of two glued $4$-simplices (decomposition of the $4$-sphere) in the Barrett-Crane model in the large internal spin regime.
\end{abstract}

\maketitle

\section{Introduction}

In spin foam models \cite{spinfoams,spinfoams2,spinfoams3,baez,spinfoam_zakopane} the asymptotic expansion of the vertex amplitude in the large spin limit is an important test for their connection to (discrete) gravity. The expansion of the known vertex amplitudes can be written in the form
\begin{equation}
 \sum_\alpha A_\alpha(\{j_i\})e^{i S_\alpha(\{j_i\})}\left(1 +O\left(|j|^{-1}\right)\right) \quad ,
\end{equation}
where $O(|j|^{-1})$ denotes more suppressed terms in the expansion. The sum over indices $\alpha$ indicates that the asymptotic expansion usually contains contributions from different sectors, which differ in their geometric and physical interpretation. According to common terminology we call $S_\alpha$ the phase or the action and $A_\alpha$ the measure factor\footnote{ Here we refer to the asymptotic measure factor of the vertex amplitude. This is not to be confused with the face and edge amplitudes (also called measure factors) of the spin foam model.}.

The value of the asymptotic expansion lies in its geometric interpretation, for example, in the case of the $6j$ symbol, the expansion consists of two terms (see for example \cite{Roberts, Hal,freidel-louapre, kaminski-steinhaus} for derivations). For both the phase is equal to the Regge action (up to a sign) associated to tetrahedra built from the spin labels. This strongly supports the claim that the Ponzano-Regge model \cite{PR,pr-model}, which utilizes ${\rm SU(2)}$ $6j$ symbols as its basic building block, is a model of $3$D quantum gravity. The modern $4$D spin foam models, such as the Barrett-Crane- \cite{barrett-crane}, EPRL- \cite{eprl} or FK-model \cite{fk}, were invented as generalizations of this case adapted to incorporate the so-called simplicity constraints. Similar results as for the $3$D model were obtained for the action of these models \cite{4dWilliams,Baez10j,freidel-louapre,Frank,Frank2,Frank3,FrankEPRL}, which is a significant support of the relation of these models to Quantum Gravity.

However, the understanding of the asymptotic expansion is far from complete: Besides the troubling fact that non-geometric configurations are contributing, even for geometric contributions the measure factors, which play an important role in asymptotic expansions, pose a challenge. From the simple example of the $6j$ symbol, one expects that they should be given in simple geometric terms. However, known spin foam amplitudes resist attempts to find such formulas. Whereas the phase contribution to the asymptotics is well known, the measure factor has remained mysterious so far.  In this paper we are considering the simplest model introduced by Barrett and Crane (BC model) in the case of two glued $4$-simplices.

It is known that this model suffers from several issues: One of them is the presence of the topological BF sector common to all spin foams \cite{Frank, freidel-louapre, Barrett2003, Baez10j}, but in addition it lacks gluing constraints \cite{Barrett1999, fk, Alesci2007}. 
The faces along which the simplices are glued have the same area yet their shapes differ in general.
The theory obtained in a semiclassical limit, called area Regge calculus \cite{Rovelli1993}, is troubled by metric discontinuities and vanishing deficit angles \cite{wainwright_williams, neiman_area}.
 For this reason the model is now commonly disregarded as a viable model for Quantum Gravity, for a recent discussion on this see \cite{bonzom_bc} and \cite{gft_bc}. This naturally raises the question why we are considering it at all.
The simple yet honest answer would be that this is the only model we can cope with, but there is a more profound reason: It is known that the geometric sector of the BC model is supported on the same configurations as the flipped Freidel-Krasnov model ( $\!\!\overline{\text{FK}}$ model in \cite{HellmannKaminski}). In the latter the $\gamma$-flatness property \cite{HellmannKaminski} takes its extreme form as complete $\text{SO}(4)$ flatness: The deficit angles on all bulk faces have to vanish like in the BC model. Recently this led to hopes that a semiclassical limit exists, for which this property can assure large scale geometricity, i.e. small curvature on small scales (see \cite{Han2013} for discussion of a double-scaling limit for the EPRL model in which the representation labels diverge, but curvature goes to zero)
\footnote{To reach this regime one needs, however, more refined  tools. For first steps in this direction using coarse-graining techniques see \cite{perfect_improve,perfect_aho, holonomy_sfm1,s3_tnw}.}.
In this context the BC model appears to be a nice `test and proving ground' \cite{Bonzom2013}. We hope that this result can provide some insight for more physical models.

In this work we mainly focus on the amplitude of the BC model associated to the geometry of two $4$--simplices glued along  all of their tetrahedra,  forming a triangulation of the $4$-dimensional sphere that is given by the product of two vertex amplitudes. The asymptotic behaviour of the latter were analysed in \cite{4dWilliams,Baez10j} and finally in \cite{freidel-louapre}. For simplicity we assume that both face and edge amplitudes are equal to one, even though their importance for the geometric interpretation has been emphasized in \cite{gft_bc}. Our result is only slightly affected by this choice. 

As a starting point for our derivation, we will use formula (125) from ref. \cite{freidel-louapre} for the $10j$ symbol, i.e. the BC amplitude of a single $4$-simplex (see \cite{kaminski-steinhaus} for the derivation of the change of measure, the normalisation factor is equal to $ \frac{ 2^4|O(4)|}{|S_3|^5}=\frac{ 2}{\pi^6}$):
\begin{equation} \label{eq:initial_formula}
 \frac{2}{\pi^6} \int_{ [0,\pi]^{10}\cap\{\tilde{G}\geq 0\}} \prod_{i<j} d\theta_{ij}\ \delta(\det \tilde{G}) \prod_{ij} \sin\theta_{ij}\ \prod_{i<j} \chi_{j_{ij}}(\theta_{ij}) \quad .
\end{equation}
Let us briefly explain the notation: Each tetrahedron of the $4$--simplex is labelled by an integer $i \in \{1,2,\dots,5\}$. Each face is shared by two tetrahedra and is hence labelled by a pair of indices $i,j$. Attached to the faces are both the spins $j_{ij}$, denoting irreducible representations of $\text{SU}(2)$, and the exterior dihedral angles $\theta_{ij}$. The function $\chi_j(\theta)$ denotes the $\text{SU}(2)$ character function \eqref{eq:character} and $\tilde{G}$ denotes the angle Gram matrix defined in appendix \ref{app:gram}.

Under the necessary assumption that $\forall_i \sum_j j_{ij}\in {\mathbb N}$, the integral \eqref{eq:initial_formula} is invariant under the action of the symmetry group ${\mathbb Z}_2^4$, which has the following mutually commuting generators of order $2$:
\begin{equation}
 g_i\colon \forall_j\ \theta_{ij}\rightarrow \pi-\theta_{ij}\ .
\end{equation}
with the relation $\prod_i g_i=1$.
Several regions of the domain of integration contribute to the asymptotics:
\begin{itemize}
 \item Certain values of $\theta_{ij}$ on the boundary of the integration domain form the degenerate sector. The contribution from this region may be even dominating in the amplitude, yet it does not correspond to a $4$-geometry, but to a so-called vector geometry or the Hodge dual of that \cite{Barrett2003}.
 \item $\theta_{ij}$ are the angles between the normals in $\mathbb{R}^4$, such that the $4$-simplex with tetrahedra perpendicular to the normals (that is determined uniquely up to a scale) has face areas $A_{ij}=2j_{ij}+1$. Given only the set of areas $\{A_{ij}\}$, there might be several configurations of angles $\{\theta_{ij}\}$ contributing.
\end{itemize}
Since we are considering two $4$-simplices, we have the same distinction for each of them. However, even if we restrict to non-degenerate configurations, the shapes (of the glued tetrahedra) do not need to match; this is a known problem of the BC model. From now on, we will denote the configurations with matching shapes as geometric and neglect all other, i.e. non-geometric, contributions to the amplitude, even though we are aware of their existence.

We will now proceed to present our result and its derivation in the next sections.

\section{Main result} \label{sec:main}

Our main result can be formulated as follows. The geometric part of the leading asymptotic expansion of the spin foam built from two glued $4$-simplices with faces $A_{ij}=2j_{ij}+1$ is given by the formula\footnote{ We integrate over all sets of lengths $l_{ij}$ that form a $4$-simplex.}
\begin{equation}\label{eq:result1}
\frac{3^3}{ 2^{23} \pi^3} \int_{{\cal C}} \prod_{i<j} dl_{ij}^2\  \prod_{i<j} \delta(A_{ij}(l)^2-A_{ij}^2)\ \frac{\prod_i {W_i'}^2}{{V'}^7}\cos^2\left(\sum_{i<j} A_{ij}(l)\theta_{ij}(l)-\frac{\pi}{4}\right) \quad ,
\end{equation}
where $W_i'$ is the  volume of $i$th tetrahedron, and $V'$ is the  volume of the $4$-simplex. The integration region $\cal C$ consists of all geometric lengths (see appendix \ref{app:length}).
The formula is valid for generic configurations $A_{ij}$ such that the map from length variables $\{l_{ij}\}\rightarrow \{A_{ij}\}$ is locally invertible for all $l_{ij}$ with given face areas constructed from them\footnote{Please note the subtlety in the notation: $A_{ij}$ denotes the area of the triangle obtained from one $4$-simplex by removing the vertices $i$ and $j$, whereas $l_{ij}$ actually is the edge connecting $i$ and $j$.}, and the reconstructed $4$-simplices are non degenerate. If the maps $\{l_{ij}\}\rightarrow \{A_{ij}\}$ are not locally invertible, this contribution to the total amplitude is less suppressed as the Hessian has an additional null eigenvector. We leave the investigation of these cases for future research.

Although this expression appears to be complicated at first sight, it is possible to perform the integration:
\begin{equation}\label{eq:result1-2}
\frac{3^3}{  2^{23} \pi^3} \sum \det \left(\frac{\partial l_{ij}^2}{\partial A_{ij}^2} \right) \frac{\prod_i {W_i'}^2}{{V'}^7}\cos^2\left(\sum_{i<j} A_{ij}(l)\theta_{ij}(l)-\frac{\pi}{4}\right) \quad ,
\end{equation}
where the summation is with respect to all possible configurations of lengths $l_{ij}$, from which the given areas can be constructed. Please note that the only non explicit factor is $\det \frac{\partial l_{ij}^2}{\partial A_{ij}^2}$, which corresponds to the change of variables.

The result in the form \eqref{eq:result1-2} can be simply translated into the geometric contribution to the asymptotic expansion for a single $4$-simplex amplitude
\begin{equation}\label{eq:result1-3}
\frac{3^{\frac{3}{2}}}{2^{\frac{23}{2}} \pi^\frac{3}{2}} \sum \pm\sqrt{\det \left(\frac{\partial l_{ij}^2}{\partial A_{ij}^2} \right)} \frac{\prod_i {W_i'}}{{V'}^\frac{7}{2}}\cos\left(\sum_{i<j} A_{ij}(l)\theta_{ij}(l)-\frac{\pi}{4}\right) \quad ,
\end{equation}
The apparent sign ambiguity needs further research.

\subsection{Comments}\label{sec:comments}

Before we discuss the derivation of this result, we would like to briefly comment on two subjects: first the choice of the examined triangulation and second the omittance of other contributing sectors.

The natural question arising is why we are considering a triangulation of the sphere and not a single $4$-simplex. In fact, the  latter can be deduced from the second version of our result \eqref{eq:result1-2} (with a sign ambiguity). Then the only non-explicit term is the Jacobian $\det \frac{\partial l^2_{ij}}{\partial A^2_{ij}}$. Surprisingly in the case of the $4$-sphere this Jacobian can be absorbed by imposing area data as constraints, such that the asymptotic formula can be simply stated in terms of length variables.

The origin of the phenomenon that the asymptotic formula for the triangulated $4$--sphere is simpler than the one for a single $4$--simplex is obscure to us. For general spin foam models it is well known that the measure factors, contrary to the phase amplitudes, are inflicted by the spread of the boundary states. Thus one would expect to find nicer results for closed triangulations. Of  course, this argument does not apply to the BC model, because its boundary space structure is trivial, i.e. the intertwiner space is $1$--dimensional and there is no choice of the spread of the boundary state.
Instead, we rather expect the imposition of the area data as constraints to be the root of this property.

The second issue we would like to address is that our result concerns only a part of the contribution to the asymptotic behaviour, yet it has a clear numerical meaning at least in the case of nondegenerate geometries. 
Precisely, if the asymptotic formula (to leading order) is given by
\begin{equation}\label{eq:form1}
 \sum_\alpha A_\alpha(\{j_i\})e^{i S_\alpha(\{j_i\})} \quad ,
\end{equation}
where all phases $S_\alpha$ are different functions of the data, then every term has a separate meaning and is interesting by itself.

In the case of the Barrett-Crane model,
the additional contributions omitted in our derivation are known to some extent. They can be treated by the method of coherent states or by the
Kirillov character formula \cite{Barrett2003}. For the case of a single simplex, their phase turns out to be zero. Thus the additional contributions to our  asymptotic formula are
indeed of the form \eqref{eq:form1} with: 
\begin{itemize}
\item $S_\alpha = \pm S_{\rm Regge}\pm S_{\rm Regge}'$ for non--matching configurations and thus differing Regge actions, 
\item $S_\alpha = 0$ for matching configurations, but opposite signs or non--geometric configurations and finally
\item $S_\alpha = \pm S_{\rm Regge}$ if one of the simplices is in a non--geometric configuration.
\end{itemize}
Our result can thus be regarded as a derivation of the measure factors for the terms in the asymptotic formula with phases $\pm 2S_{\rm Regge}$.

Before we discuss this result further, we will briefly present its derivation in the following two sections.

\section{Derivation} \label{sec:derivation}

In this section we sketch our proof, leaving technical details to the next section. In order to simplify notation and to avoid keeping track of numerical factors, we absorb the combinatorial factors into the definition of the volume of simplices, e.g. $V$ is $4!$ times the volume of a $4$-simplex and $W_i$ is $3!$ times the volume of the $i$th tetrahedron, $A_{ij}$ denotes the area of the face. Faces are labelled by two vertices, i.e. those which do not belong to the face.
From now on, we use the convention that
$\vec{l}$, $\vec{A}$, $\vec{W}$ denote vectors consisting of all lengths, areas and tetrahedra respectively. For any vector $\vec{m}$,  $|m|$ denotes its Cartesian length. Since the dihedral angles of a $4$-simplex sit on the faces of the simplex, i.e. the triangles, they carry the same labels as the faces.

\subsection{Stationary point conditions}

Starting from \eqref{eq:initial_formula}, we use the character formula ($\theta$ is the ${\rm SU}(2)$ angle)
\begin{equation}\label{eq:character}
\chi_j(\theta)=\frac{\sin(2j+1)\theta}{\sin\theta}=\frac{e^{i(2j+1)\theta}-e^{-i(2j+1)\theta}}{2i\sin\theta}
\end{equation}
to divide the integral into several parts (the sines from \eqref{eq:initial_formula} cancel with the sines from the integral)
\begin{equation} \label{eq:initial_action}
\frac{\pm 1}{2^{10}} \frac{ 2 }{\pi^6} \int_{[0,\pi]^{10}} \prod_{i<j} d\theta_{ij}\ \delta(\det \tilde{G})\  e^{i\sum \pm (2j_{ij}+1)\theta_{ij}}\\
 =\pm\frac{|A|}{2^{10}\pi^7} \int_{[0,\pi]^{10}} \prod_{i<j} d\theta_{ij}\int d\rho  e^{i\left(\sum \pm A_{ij}\theta_{ij}-|A|\rho \det \tilde{G}\right)}
\end{equation}
to which stationary point method can be applied -- boundary terms contribute non geometric asymptotics as shown in \cite{freidel-louapre} and will be omitted. The stationary phase conditions are:
\begin{itemize}
\item For $\frac{\partial S}{\partial \theta_{ij}} = 0$:
\begin{equation} \label{eq:stat1}
 \pm A_{ij}=\lambda \frac{\partial \det \tilde{G}}{\partial\theta_{ij}} \quad ,
\end{equation}
where we introduced notation $\lambda := |A|\rho$.
\item For $\frac{\partial S}{\partial \rho} = 0$:
\begin{equation} \label{eq:stat2}
 \det \tilde{G}=0 \quad .
\end{equation}
\end{itemize}
The latter condition ensures that $\theta_{ij}$ are geometric angles. In fact, the vanishing of the determinant of the angle Gram matrix ensures that the flat $4$-simplex with the given angles between normals (associated to tetrahedra) exists. This simplex is determined uniquely up to rotations, parity transformations and scaling. However these normals are not necessarily outward pointing. The before mentioned generators of the ${\mathbb Z}_2^4$ symmetry correspond to the change of normals
\footnote{ Notice that $\prod  \tilde{g}_i$ acts as an element of $O(4)$ and vectors $n_i$ are determined only up to $O(4)$ transformations.}
\begin{equation}
 \tilde{g}_i\colon n_i\rightarrow -n_i \quad .
\end{equation}
Using this symmetry we can restrict our consideration to the case where all normals are outward pointing. The contribution from the other stationary points are the same and are taken into account by multiplication of the result by $2^4$.

Under the restriction that the normals to the tetrahedra should be outward pointing, every $4$--simplex determined by the normals
(that is unique up to a scale) satisfies:
\begin{equation}
 A'_{ij}=\lambda'\frac{\partial \det \tilde{G}}{\partial\theta_{ij}},\quad \lambda'=-\frac{\prod W_i^2}{4 V^7} \quad .
\end{equation}
The details of the derivation are similar to \cite{kaminski-steinhaus} and are also briefly discussed in section \ref{sec:technical}. It turns out \cite{freidel-louapre} that under such restriction only two of the integrals actually have  a point of stationary phase, namely if all signs are either `$+$' or `$-$'; then the action in \eqref{eq:initial_action} is the first order Regge action \cite{Barrett:1994nn}. Their respective contributions are related by complex conjugation, such that the final result is purely real. Hence, we will consider only the `$+$' sign case. The calculation for the `$-$' sign case works analogously.
In this case the stationary point conditions \eqref{eq:stat1} and \eqref{eq:stat2} are solved by
the angles $\theta_{ij}$ between outward pointing normals to the $4$-simplex with face areas  $\{A_{ij}\}$ and the Lagrange multiplier $\lambda'$.

\subsection{The hessian}

Most of the calculations following in the remaining sections of this work were derived in great detail for the $3$D case \cite{kaminski-steinhaus} and can be performed almost analogously in the situation under discussion. Even though the calculations are presented in a self-consistent way, we highly recommend that interested readers also study \cite{kaminski-steinhaus} to better understand the crucial ideas and background information of this approach.

The matrix of second derivatives has the form 
\begin{equation}
\mathcal{H} := -i|A|\left(\begin{array}{cc}
        0 &\frac{\partial\det\tilde{G}}{\partial\theta_{ij}}\\
    \frac{\partial\det\tilde{G}}{\partial\theta_{km}} & \rho
\frac{\partial\det\tilde{G}}{\partial\theta_{ij}\partial\theta_{km}}\\
       \end{array}
\right) 
\end{equation}
and its inverse is given by ($c$ is a constant)
\begin{equation}
 \mathcal{H}^{-1}=i\left(\begin{array}{cc}
        \frac{c}{|A|^2} &\frac{1}{|A|}\frac{\partial\lambda}{\partial A_{ij}}\\
   \frac{1}{|A|}\frac{\partial\lambda}{\partial A_{kl}} & \frac{\partial\theta_{ij}}{\partial
A_{kl}}\\
       \end{array}
\right) \ .
\end{equation}
We know that $\frac{\partial\theta_{ij}}{\partial A_{kl}}$ is symmetric and has exactly one null eigenvector $\vec{A}$, since given the dihedral angles, the $4$-simplex is determined up to scale: We can scale all edges by $\xi$ (then all areas are scaled by $\xi^2$) to obtain a $4$-simplex with the same dihedral angles. In fact, this is the only  remaining freedom, once all $\theta_{ij}$ are fixed.
We can write the matrix in the basis with $\frac{\vec{A}}{|A|}$ as a basis vector
\begin{equation}
 i\left(\begin{array}{cccc}
        \frac{c}{|A|^2} &\frac{1}{|A|}\frac{\partial\lambda}{\partial A}&\cdots&\cdots\\
    \frac{1}{|A|}\frac{\partial\lambda}{\partial A}&0&0&0\\
    \vdots& 0&\frac{\partial\theta_{ij}}{\partial
A_{kl}}&\vdots\\
   \vdots&0&\cdots&\ddots
       \end{array}
\right) \quad ,
\end{equation}
where $\frac{\partial}{\partial A}:=\frac{A_{ij}}{|A|}\frac{\partial}{\partial A_{ij}}$\footnote{We are using Einstein's summing convention for the index pair $(ij)$ with $i \leq j$.}.
The determinant of $(-\mathcal{H}^{-1})$ is thus equal to
\begin{equation}
 \det (-\mathcal{H}^{-1})=-(-i)^{11}\Big(\,\underbrace{\frac{1}{|A|}\frac{\partial\lambda}{\partial
A}}_{= \frac{\lambda}{ |A|^2}} \, \Big)^2\ {\det}'\frac{\partial\theta_{ij}}{\partial
A_{kl}} \quad ,
\end{equation}
where ${\det}' \frac{\partial \theta_{ij}}{\partial A_{kl}}$ is the determinant of the matrix $\frac{\partial \theta_{ij}}{\partial A_{kl}}$ restricted to the subspace orthogonal to the vector $\vec{A}$ (the only null eigenvector of this symmetric matrix)\footnote{Consider a symmetric matrix $M$ with exactly one null eigenvector. The determinant of this matrix restricted to the subspace orthogonal to that eigenvector is given by ${\det}'M:= \sum_i M^*_{ii}$, where $M^*_{ii}$ denotes the $(i,i)$th minor of the matrix $M$. See \cite{kaminski-steinhaus} for more details.}.
Now we will expand
\begin{equation}\label{matrix-tlA}
 \frac{\partial \theta_{ij}}{\partial A_{kl}} = \frac{\partial \theta_{ij}}{\partial l_{mn}} \frac{\partial l_{mn}}{\partial A_{kl}} \quad .
\end{equation}
Scaling symmetry (see appendix \ref{app:scaling}) ensures that $A_{kl} \frac{\partial l_{ij}}{\partial A_{kl}} = \frac{1}{2} l_{ij}$. A similar argument of scaling shows that $\vec{l}$ is a null eigenvector of $\left(\frac{\partial \theta_{ij}}{\partial l_{kl}}\right)$.  As before, we write the matrix $\frac{\partial l_{ij}}{\partial A_{kl}}$ in the adapted basis (with the  normalized vectors $\frac{\vec{l}}{|l|}$ and $\frac{\vec{A}}{|A|}$)
\begin{equation}
 \left(\begin{array}{c|ccc}
        \frac{|l|}{ 2 |A|} & 0 &\cdots & 0\\
        \hline
    0& & & \\
    \vdots & &\frac{\partial l_{ij}}{\partial
A_{kl}}&  \\
0 & & & 
       \end{array}
\right) \quad .
\end{equation}
Notice moreover that\footnote{The first relation is the well-known Schl\"afli identity.}
\begin{equation}
 A_{ij}\frac{\partial\theta_{ij}}{\partial
l_{kl}}=l_{kl}\frac{\partial\theta_{ij}}{\partial
l_{kl}}=0 \quad ,
\end{equation}
thus working in the bases with vectors $\frac{\vec{l}}{|l|}$ and $\frac{\vec{A}}{|A|}$ we can prove 
(using \eqref{matrix-tlA}) that
\begin{equation}
 {\det}'\frac{\partial\theta_{ij}}{\partial
A_{kl}}={\det}'\frac{\partial\theta_{ij}}{\partial
l_{kl}}\ \frac{ 2 |A|}{|l|}\ {\det}\frac{\partial l_{ij}}{\partial
A_{kl}} \quad .
\end{equation}
Together with the formula
\begin{equation}
 {\det}'\frac{\partial\theta_{ij}}{\partial
l_{kl}}= 2^{-10} \frac{|A||l|}{\prod W_i^2}V^7\frac{\prod l_{ij}}{\prod A_{ij}} \quad ,
\end{equation}
which is calculated in section \ref{sec:technical}, we get
\begin{equation} \label{eq:det-Hessian}
 \det (-\mathcal{H}^{-1})=-i \, 2^{ -9}  (\prod W_i^2)\frac{\prod l_{ij}}{\prod A_{ij}} V^{-7} \frac{1}{|A|^2} \det \frac{\partial l_{ij}}{\partial A_{kl}} \quad .
\end{equation}
The combined contribution from two (conjugated) stationary points for a single $4$--simplex is thus
\begin{equation}
 \pm \frac{|A|}{ 2^{\frac{7}{2}} \pi^\frac{3}{2}}{\sqrt{|\det (-\mathcal{H}^{-1})|}}\cos\left(\sum A_{ij}\theta_{ij}-\frac{\pi}{4}\right) \quad .
\end{equation}
Inserting the determinant of the Hessian matrix \eqref{eq:det-Hessian}, one obtains the result \eqref{eq:result1-3} with a sign ambiguity and the Jacobian $\frac{\partial l_{ij}}{\partial A_{ij}}$ as the only implicit term.

Precisely this Jacobian can be `absorbed' in the case of two $4$--simplices, glued together via all of their tetrahedra, i.e. a triangulation of the $4$--sphere. However this only works if the shapes of the two $4$--simplices match; we call this a geometric configuration.
This contribution can be written as a sum over the stationary points for one integral of the square of the amplitude. We obtain
\begin{equation}
 \sum \frac{|A|^2|\det (-\mathcal{H}^{-1})|}{ 2^{7} \pi^3}\cos^2\left(\sum A_{ij}\theta_{ij}-\frac{\pi}{4}\right) \quad ,
\end{equation}
where the summation is over all angles for which the areas $A_{ij}$ match. We rewrite this summation by an integral over the edge lengths (squared), where we impose the area matching as a constraint:
\begin{equation}
 \int \prod dl_{ij}^2 \frac{\prod A_{ij}}{\prod l_{ij}} \det \frac{\partial A_{ij}}{\partial l_{kl}} \prod \delta(A_{ij}(l)^2-A_{ij}^2) \quad .
\end{equation}
Thus the result, after counting also the contributions from stationary points related by the ${\mathbb Z}_2^4$ symmetry, is
\begin{equation}
\frac{1}{ 2^{12} \pi^3} \int \prod dl_{ij}^2  \prod \delta(A_{ij}(l)^2-A_{ij}^2) \frac{\prod W_i^2}{V^7}\cos^2\left(\sum A_{ij}\theta_{ij}-\frac{\pi}{4}\right) \quad .
\end{equation}

In order to arrive at the result \eqref{eq:result1}, one just has to extract the combinatorial factors out of $V$ and $W_i$ as described at the beginning of section \ref{sec:derivation}.
In the following section we will provide all the necessary technical details to fill the gaps on the presented derivation. An even more thorough discussion can be found in \cite{kaminski-steinhaus}.

\section{Technical details} \label{sec:technical}

\subsection{General facts}

Let us also remind some general facts (we assume that $\{A_{ij}\}$ locally constitute a good coordinate system)
\begin{align}
&A_{ij}\frac{\partial\theta_{ij}}{\partial
l_{kl}}=l_{kl}\frac{\partial\theta_{ij}}{\partial
l_{kl}}=0\ ,\  
A_{ij}\frac{\partial\theta_{ij}}{\partial
A_{kl}}=A_{kl}\frac{\partial\theta_{ij}}{\partial
A_{kl}}=0\ , \ 
\frac{\partial\theta_{ij}}{\partial
A_{kl}}=\frac{\partial\theta_{kl}}{\partial
A_{ij}} \ ,\\
&A_{ij}=\lambda\frac{\partial\det\tilde{G}}{\partial\theta_{ij}}\ ,\ \lambda= -\frac{\prod
W_i^2}{4V^7} \ ,\ 2A_{ij} V=W_iW_j\sin\theta_{ij}\ ,
\end{align}
which can be proven analogously as in \cite{kaminski-steinhaus}.
For the flat $4$-simplex holds
\begin{align}
&\frac{\partial\lambda}{\partial A_{ij}}
\frac{\partial\det\tilde{G}}{\partial\theta_{kl}}+
\frac{\partial\theta_{ij}}{\partial
A_{mn}}\lambda\frac{\partial^2\det\tilde{G}}{\partial\theta_{mn}\partial\theta_{kl}}=
\delta_{(ij),(kl)} \quad ,\\
&\exists_c\
\frac{\partial\det\tilde{G}}{\partial\theta_{kl}}c+\lambda\frac{\partial\lambda}{\partial
A_{ij}}\frac{\partial^2\det\tilde{G}}{\partial\theta_{ij}\partial\theta_{kl}}=0 \quad ,\\
&\frac{\partial\lambda}{\partial
A_{mn}}\frac{\partial\det\tilde{G}}{\partial\theta_{mn}}=1 \quad .
\end{align}
Interested reader should consult appendix C in \cite{kaminski-steinhaus} for more details.

\subsection{Limits from curved simplex}

In the following two subsections we will consider a curved $4$-simplex with edge lengths $\epsilon \, l_{ij}$ from which we will derive several quantities for the flat simplex. To do so we will heavily rely on the duality relations of length and angle Gram matrices in the curved case, which are nicely presented in \cite{Kokkendorff}. The desired quantities will be obtained in the limit $\epsilon\rightarrow 0$.
For any dimension $d$ holds
\begin{align}
 \det G^{(d)}=\epsilon^{2d}(V^{(d)})^2+O(\epsilon^{2d+2}) \quad ,
\end{align}
where $G^{(d)}$ is the length Gram matrix and $V^{(d)}$ is $d!$ times the volume of $d$ dimensional simplex
\footnote{Please note that our convention affects only tetrahedra and $4$-simplices, but not areas and edge lengths.}. We denote 
\begin{equation}
 V^{(3)}=W,\quad V^{(4)}=V,\quad  V^{(2)}=2A .
\end{equation}

Using the results of \cite{Kokkendorff} we arrive at the following identity for $\det' \tilde{G}$ (for $d = n - 1 =4$)
\begin{equation}
{\det}'\tilde{G}^{\epsilon}=\sum_i \tilde{G}^*_{ii}=\left(\sum_i G_{ii}^*\right)\frac{(\det
G)^{3}}{\prod G_{ii}^*}
=\left(\epsilon^{6} \sum_i W_i^2\right)\frac{ \epsilon^{24} \,
V^6}{\epsilon^{30} \, \prod W_i^2}+O(\epsilon^2)=|W|^2\frac{V^6}{\prod W_i^2}+O(\epsilon^2) \quad .
\end{equation}
and, moreover, also using \cite{Kokkendorff}, we have
\begin{equation}
 \det \tilde{G}^{\epsilon}=\frac{(\det G)^4}{\prod \det G_i}=-\frac{\left(\epsilon^8 V^2\right)^4}{\prod \epsilon^6 W_i^2}+O(\epsilon^4)=\epsilon^2\frac{V^8}{\prod W_i^2}+O(\epsilon^4) \quad .
\end{equation}
We will use the latter identity in the following formula for a curved $4$-simplex, which has been stated in \cite{DF} (for derivation see \cite{kaminski-steinhaus}):
\begin{equation}
 \det \frac{\partial\theta^\epsilon_{ij}}{\partial l^\epsilon_{km}}=(-1)
\frac{\prod\sin
l^\epsilon_{ij}}{\prod\sin\theta^\epsilon_{ij}}\left(\frac{\det\tilde{G}}{\det G}\right)^{3}\ ,
\end{equation}
where $l^\epsilon_{ij}=\epsilon l_{ij}$. Thus
\begin{equation}
 \epsilon^{-10}\det \frac{\partial\theta^\epsilon_{ij}}{\partial l_{km}}=\det \frac{\partial\theta^\epsilon_{ij}}{\partial l^\epsilon_{km}}=-\epsilon^{10}\frac{\prod l_{ij}}{\prod \sin\theta_{ij}}
\frac{\epsilon^6\frac{V^{24}}{\prod W_i^6}}{\epsilon^{24} V^{6}}+O(\epsilon^{-6})
\end{equation}
and finally substituting $\sin \theta_{ij}=\frac{2A_{ij}V}{W_iW_j}$
\begin{equation}
 \det \frac{\partial\theta^\epsilon_{ij}}{\partial l_{km}}=-\frac{\epsilon^2}{2^{10}} \frac{\prod l_{ij}}{\prod A_{ij}} \frac{V^8}{\prod W_i^2}
 +O(\epsilon^4) \quad .
\end{equation}

\subsection{Limits of determinant with one null eigenvector}

If the matrix $M^\eta$ has in the limit  $\eta \rightarrow 0$ one  null eigenvector $\vec{m}$ and is holomorphic in $\eta$ then the following expansion around $\eta=0$ holds:
\begin{equation}
 \det M^\eta={\det}' M \frac{(\vec{m},M^\eta\vec{m})}{|m|^2}+O(\eta^2) \quad .
\end{equation}
This expansion can be immediately applied to $\det \tilde{G}$, since $\tilde{G}$ (and thus also  $\det \tilde{G}$) is holomorphic in $\epsilon^2$ and $\tilde{G}$ has one null eigenvector $\vec{W}$ in the flat case
\begin{equation}
 \det \tilde{G}^{\epsilon}= {\det}'\tilde{G} \frac{(\vec{W},\tilde{G}^\epsilon\vec{W})}{|W|^2}+O(\epsilon^4) \quad .
\end{equation}
Furthermore $(\vec{W},\tilde{G}^\epsilon\vec{W})=O(\epsilon^2)$ and can be expanded as follows:
\begin{equation}
 2(\vec{W},\tilde{G}^\epsilon\vec{W})=\epsilon\frac{\partial}{\partial \epsilon}(\vec{W},\tilde{G}^\epsilon\vec{W})+O(\epsilon^3)=(\vec{W},l_{ij}\frac{\partial \tilde{G}^\epsilon}{\partial l_{ij}}\vec{W})+O(\epsilon^3) \quad ,
\end{equation}
and 
\begin{equation}
(\vec{W},l_{ij}\frac{\partial \tilde{G}^\epsilon}{\partial l_{ij}}\vec{W})
=l_{ij}W_k W_l \sin \theta^\epsilon_{kl} \frac{\partial \theta^\epsilon_{kl}}{\partial l_{ij}}
=2Vl_{ij}A_{kl}\frac{\partial \theta^\epsilon_{kl}}{\partial l_{ij}}+O(\epsilon^3) \quad .
\end{equation}
Eventually we can summarize these results in a concise form:
\begin{equation}
 l_{ij}A_{kl}\frac{\partial \theta^\epsilon_{kl}}{\partial l_{ij}}= 
 \frac{|W|^2}{ V}\frac{\det \tilde{G}^\epsilon}{{\det}'\tilde{G}}+O(\epsilon^3)=
 \epsilon^2 V +O(\epsilon^3) \quad .
\end{equation}
On the other hand, we can apply the same argument to $\frac{\partial\theta^\epsilon_{ij}}{\partial l_{km}}$, since it also is holomorphic in $\epsilon^2$ and has exactly one null eigenvector in the flat case, namely $\vec{l}$ and respectively $\vec{A}$ (it is not symmetric). Thus
\begin{equation}
 \det \frac{\partial\theta^\epsilon_{ij}}{\partial l_{km}}=
 {\det}'\frac{\partial\theta_{ij}}{\partial l_{km}}\frac{\left(\vec{A},\frac{\partial\theta^\epsilon_{ij}}{\partial l_{km}}\vec{l}\right)}{|A||l|}+O(\epsilon^3)
\end{equation}
Substituting $\det \frac{\partial\theta^\epsilon_{ij}}{\partial l_{km}}$ and $\left(\vec{A},\frac{\partial\theta^\epsilon_{ij}}{\partial l_{km}}\vec{l}\right)=l_{ij}A_{kl}\frac{\partial \theta^\epsilon_{kl}}{\partial l_{ij}}$ we obtain
\begin{equation}
 {\det}'\frac{\partial\theta_{ij}}{\partial l_{km}}= 2^{-10} \frac{|A||l|}{\prod W_i^2}V^7\frac{\prod l_{ij}}{\prod A_{ij}} \quad .
\end{equation}

\section{Conclusions}\label{sec:conclusions}

In this work we have derived a first asymptotic formula for the Barrett-Crane spin foam model, expressed completely in terms of simple geometric quantities, which sheds new light on the possible asymptotic behaviour of more complicated spin foam models. In fact it suggests that edge lengths are the more suitable variables to be used in the asymptotic limit than area variables. Exactly this might be a root of the problem to derive a measure factor in the asymptotic limit of modern spin foam models. 

Let us put our derivation in a geometric but heuristic context. Expanding the $\text{SU}(2)$ characters in \eqref{eq:initial_formula} into exponentials one can obtain the so-called bare action for arbitrary foams. Putting aside  the troubling fact that we sum over spins instead of integrating, we can regard the models as a Feynman path integral. The action turns out to be the action of first order (area) Regge calculus, where both areas and 4D dihedral angles are independent variables. 
One can then argue that the semiclassical expansion of the integral is governed by the expansion of this action
around the solutions to the classical equations of motion, which differ from standard Regge calculus \cite{Barrett1999,Dittrich:2008va,wainwright_williams, regge_new_angle}. If we restrict ourselves to geometric solutions for which the shapes of tetrahedra match then there are only flat solutions, but with arbitrary non-matching orientations. 
Flatness is imposed by the equations of motion  originating from the variation over areas. In our asymptotic analysis, however, the areas are fixed and uniformly large.  As the suppression of non-flat solutions is related to summing (integrating) over spins (see \cite{HellmannKaminski, Bonzom}), and we do not sum over spins, this part of the equations of motion is absent in our asymptotics.  In our case non-flat solutions are not suppressed.

The results of \cite{Bonzom} show that the modern spin foam models correspond to (area, coherent states)- Regge calculus. This is yet another version of Regge calculus, also leading to the flatness problem, yet with matching area shapes \cite{Bonzom}. It is not clear how to extend our result in this setting, nevertheless it suggests
new types of factors that might appear in the asymptotic expansion, as the just derived formula can also be stated for a single $4$-simplex, although with an additional factor $\sqrt{\det\frac{\partial A_{ij}}{\partial l_{ij}}}$.

It is also worth to speculate about the suggestions from our work for measure factors \cite{Hamber:1997ut} in (linearized) Regge calculus due to its close relation to spin foams in the asymptotic limit. One of the typical constraints for such measures derived in \cite{Dittrich:2011vz} is triangulation independence, i.e. invariance under Pachner moves. 
It can be shown (see \cite{Bianca-Sebastian-Wojtek}) that a product (or fraction) of simple geometric quantities (like volumes, areas or edge lengths) cannot be invariant. However, if the non-local change of variables plays a role in larger simplicial complexes, then the invariance might be restored.

Finally, in the case when the map $\{l_{ij}\}\rightarrow \{A_{ij}\}$ is not locally invertible, the Hessian has an additional null eigenvector and the asymptotic contribution is less suppressed. If a similar phenomenon exists in the Lorentzian EPRL--FK model it would directly influence the result of \cite{Aldo}, which is derived under the assumption of a certain scaling behaviour of the measure factor. Our result suggests, which configurations might show such an anomalous scaling behaviour and are thus crucial for research in this direction.

As we have shown in this work there still exist new insights that can be drawn from the Barrett-Crane model. 
Let us notice, that the derivation is restricted only to a very specific example and it is not clear how to extend it to more general situation. We hope however that our methods presented here and in \cite{kaminski-steinhaus} can be expanded to the more advanced spin foam models and might shed some new light on the pressing questions such as the measure factor, the flatness issue or the suitable choice of variables and semi-classical limits.

\begin{acknowledgments}
We thank Bianca Dittrich, Frank Hellmann, Jerzy Lewandowski, Hal Haggard and Ryszard Kostecki for fruitful discussions and constant encouragements to write this paper. We would also like to thank Daniele Oriti and the referees for their extensive comments on the previous version of this paper. W.K. acknowledges partial support by the grant ``Maestro'' of  Polish Narodowe Centrum Nauki nr 2011/02/A/ST2/00300 and the grant of Polish Narodowe Centrum Nauki number 2012/05/E/ST2/03308. S.St. gratefully acknowledges support by the DAAD (German Academic Exchange Service) and would like to thank Perimeter Institute for an Isaac Newton Chair Graduate Research Scholarship. This research was supported in part by Perimeter Institute for Theoretical Physics. Research at Perimeter Institute is supported by the Government of Canada through Industry Canada and by the Province of Ontario through the Ministry of Research and Innovation.
\end{acknowledgments}

\begin{appendix}
\section{Auxiliary definitions and computations} \label{sec:appendices}

\subsection{Definition of the Gram matrix} \label{app:gram}

For the set of numbers $\{\theta_{ij}\in(0,\pi)\}$,  the angle Gram matrix $\tilde{G}$ is defined as 
\begin{equation}
 \tilde{G}=[\cos\theta_{ij}]=\left(\begin{array}{cccc}
            1 &\cos\theta_{12}&\cdots &\cos\theta_{15}\\
            \cos\theta_{12}&1&\cdots &\cos\theta_{25}\\
            \vdots &\vdots &\ddots &\vdots\\
            \cos\theta_{15}&\cos\theta_{25}&\cdots&1
           \end{array}\right)\quad ,
\end{equation}
with the convention that $\theta_{ii}=0$. It is well--known that
for $\{\theta_{ij}\}$ satisfying $\det\tilde{G}=0$ ($\tilde{G}$ is semi--positive definite with exactly one null eigenvector) we can associate normals $n_i$ to the tetrahedra forming a $4$--simplex, such that $n_i\cdot n_j=\cos\theta_{ij}$ (see for example \cite{kaminski-steinhaus}). In the case when these normals are all outward (or inward) pointing we will call $\theta_{ij}$ an exterior dihedral angle. In such a case the $4$--simplex is uniquely determined up to a scale and $\tilde{G}$ has one null eigenvector $(W_1,\cdots, W_5)$. Using this fact we can compute
\begin{equation}
 \frac{\partial \det \tilde{G}}{\partial\theta_{ij}}=-2{\det}' \tilde{G} \frac{W_iW_j\sin\theta_{ij}}{|W|^2} \quad ,
\end{equation}
but  $2A_{ij}V=W_iW_j\sin\theta_{ij}$ and thus
\begin{equation}
 \frac{\partial \det \tilde{G}}{\partial\theta_{ij}}=
 \underbrace{-\frac{2{\det}' \tilde{G}}{|W|^2}V }_{\lambda^{-1}} A_{ij}
\end{equation}
but we also have (see also \cite{kaminski-steinhaus})
\begin{equation}
{\det}'\tilde{G}=\left(\sum_i W_i^2\right)\frac{
V^6}{\prod W_i^2} \quad .
\end{equation}
Thus we conclude
$\lambda=-\frac{\prod W_i^2}{4V^7}$.

\subsection{Geometric values of $l_{ij}$} \label{app:length}

In any dimension $d$ in order for the set of numbers $\{l_{ij}\}\in [0,\infty]^{\frac{(d+1)d}{2}}$ to be a set of lengths of a $d$-simplex the following Menger condition \cite{menger} must hold: The Cayley- Menger matrix
\begin{equation}
 C=\left(
 \begin{array}{ccccc}
  0 &1&1 &\cdots &1\\
  1 &0&l_{12}^2&\cdots &l_{1(d+1)}^2\\
  1 & l_{12}^2 &0 &\cdots& l_{2(d+1)}^2\\
  \vdots &\vdots &\vdots&\ddots &\vdots\\
  1 &l_{1(d+1)}^2&l_{2(d+1)}^2  &\cdots  &0
 \end{array}
\right)
\end{equation}
must have $1$ positive and $d+1$ negative eigenvalues. This is equivalent to the following conditions on the determinants of the upper left corner $n\times n$ submatrices $C_{n}$ of the matrix $C$:
\begin{equation}
 \forall_{2\leq n\leq d+1}\colon\ (-1)^n\det C_n<0 \quad .
\end{equation}
We denote the closure of the region of such $l_{ij}$ by $\cal C$.

\subsection{Scaling} \label{app:scaling}

By applying the field $l_{ij}\frac{\partial}{\partial l_{ij}}$ on any function of $\{l\}$, we can deduce the scaling of the respective function, e.g. the scaling of areas:
\begin{equation}
 l_{ij}\frac{\partial}{\partial l_{ij}}=l_{ij}\frac{\partial A_{kl}}{\partial l_{ij}}\frac{\partial}{\partial A_{kl}} \quad ,
\end{equation}
but since $l_{ij}\frac{\partial A_{kl}}{\partial l_{ij}}=2A_{kl}$ (areas are two dimensional), it follows:
\begin{equation}
 A_{kl}\frac{\partial}{\partial A_{kl}}=\frac{1}{2}
 l_{ij}\frac{\partial}{\partial l_{ij}} \quad .
\end{equation}
\end{appendix}

\bibliography{bc-ref}

\end{document}